\documentclass[12pt]{article}
\usepackage{amssymb}
\def\inh{\vskip 0.075truein \noindent\hangindent=12 pt \hangafter=1}

\usepackage{amsthm}\theoremstyle{remark}

\newcommand{\bte}{\begin{quote}\begin{theorem}}
\newcommand{\ete}[1]{\label{#1}\end{theorem}\end{quote}}
\newcommand{\bcom}{\begin{quote}\end{quote}}
\newcommand{\bex}{\begin{quote}\begin{example}}
\newcommand{\eex}[1]{\label{#1}\end{example}\end{quote}}
\newcommand{\bcon}{\begin{quote}\begin{conclusion}}
\newcommand{\econ}[1]{\label{#1}\end{conclusion}\end{quote}}
\newcommand{\bdefi}{\begin{quote}\begin{definition}}
\newcommand{\edefi}[1]{\label{#1}\end{definition}\end{quote}}

\newcommand{\blem}{\begin{quote}\begin{lemma}}
\newcommand{\elem}[1]{\label{#1}\end{lemma}\end{quote}}

\newcommand{\bpr}{\begin{quote}\begin{problem}}
\newcommand{\epr}[1]{\label{#1}\end{problem}\end{quote}}

\newcommand{\f}{\frac}
\newcommand{\p}{\partial}
\newcommand{\n}{\nonumber \\}
\newcommand{\inti}{\int_{-\infty}^\infty}
\newcommand{\beq}{\begin{eqnarray}}
\newcommand{\eeq}[1]{\label{#1}\end{eqnarray}}
\newcommand\eq[1]{(\ref{#1})}
\newcommand{\bfi}{\begin{figure}[24]}
\newcommand{\efi}[1]{\caption{\label{#1}}\end{figure}}
\newcommand\fig[1]{Fig.~\ref{#1}}
\newcommand{\res}{respectively}
\newcommand\gl{\left}
\newcommand\gr{\right}
\newcommand{\bfm}[1]{\mbox{\boldmath ${#1}$}}

\newcommand{\CG}{{\cal G}}

\newcommand{\CM}{{\cal M}}

\newcommand{\CQ}{{\cal Q}}

\newcommand{\Ga}{\alpha}
\newcommand{\Gb}{\beta}
\newcommand{\Gd}{\delta}

\newcommand{\Gf}{\phi}

\newcommand{\Gg}{\gamma}

\newcommand{\Gk}{\varkappa}
\newcommand{\Gl}{\lambda}

\newcommand{\Gm}{\mu}

\newcommand{\Gt}{\theta}

\newcommand{\Gr}{\varrho}

\newcommand{\Gs}{\sigma}

\newcommand{\Go}{\omega}

\newcommand{\GD}{\Delta}
\newcommand{\GF}{\Phi}

\newcommand{\az}[1]{Sect.$\!$ \ref{#1}}
\newcommand\D{\,\mathrm{d}}
\newcommand\I{\mathrm{i}}
\newcommand\E{\mathrm{e}}
\newcommand{\bexe}{\begin{quote}\begin{exercise}\inh}
\newcommand{\eexe}[1]{\label{#1}\end{exercise}\end{quote}}

\usepackage{graphics,graphicx}
\usepackage{color}

\begin{document}
{\large
\title{Structural discontinuity as generalized strain and Fourier transform for discrete-continuous systems}
\author{ Leonid I. Slepyan}
\date{\small{{\em School of Mechanical Engineering, Tel Aviv University\\
P.O. Box 39040, Ramat Aviv 69978 Tel Aviv, Israel}
}}
\maketitle

\vspace{-4mm}\noindent
{\small Email: leonid@eng.tau.ac.il (L.I.Slepyan)}

\vspace{4mm}\noindent
{\bf Abstract}
We consider a segmented structure, possibly connected with a continuous medium, as initially homogeneous, where discontinuities arise as localized strains induced by self-equilibrated localized actions. Under this formulation augmented by interface conditions, the linearized formulation remains valid. This approach eliminates the need for examining separate sections with subsequent conjugation. Only conditions related to the discontinuities should be satisfied, while the continuity in other respects preserves itself automatically. No obstacle remains for the continuous Fourier transform. For a uniform partitioning, the discrete transform is used together with the continuous one. We demonstrate the technique by obtaining the Floquet wave dispersive relations with their dependence upon interface stiffness. To this end, we briefly consider the flexural wave in the segmented beam on Winkler's foundation,  the gravity wave in a plate (also segmented) on deep water and the Floquet-Rayleigh wave in such a plate on an elastic half-space. Besides, we present the wave equations developed for an elastic medium with discontinuities.

\section{Introduction}
In this paper, we show how the continuous Fourier transform, an effective method in the linear analysis commonly used for a system homogeneous in the respective coordinate, can be extended to a segmented structure possibly interacting with a continuous medium.

Let a homogeneous system have point contacts with some other structures. For example, it could be an elastic beam supported by deformable rods.  We can carry out the Fourier transform with accounting for the (unknown) contact forces and obtaining then the latter based on respective force-displacement relations. Note that in the case of a distributed contact, we end up with an integral equation on the corresponding domain. In the case of discrete supports distributed uniformly, the discrete Fourier transform leads to a single ratio. Gueorguiev et al. (2000) examined such a system as a waveguide with several periodical arrays of attachments.

It may seem less evident that similar considerations are applicable not only for the contact inhomogeneities but local structural discontinuities as well. Note that structures with discontinuities present in different areas. An array of surface cracks in a material (in this connection, see Joglekar and Mitran, 2016) and a cracked sea ice cover can serve as examples. Segmented constructions can also be mentioned such as stationary or pontoon bridges. Note that in the ice cover case, in-plane compressive forces and the crack closure effect strongly affect the `effective' hinge stiffness manifested in angle interactions of the adjacent sections. In this connection, see Rice and Levy, 1972; Dempsey et al., 1995; Slepyan et al., 1995. (The in-plane forces also play a nontrivial role in the analysis developed below, see \az{lsel}.)

We consider an analytical technique that permits the Fourier integral transform in studying the static deformations, oscillations, and waves in discrete-continuous structures. As illustrative examples, we examine a homogeneous elastic beam (or a plate) divided into sections and coupled with a continuous medium. We show two possible configurations schematically in \fig{fin} (the parts can rotate and shift transversely relative to each other under the bending moment and transverse force, \res).

\vspace*{-15mm}

\begin{figure}[h!] 

\vspace{-0mm}\hspace{-15mm}
\hspace{30mm}\includegraphics*[width=0.8\textwidth]{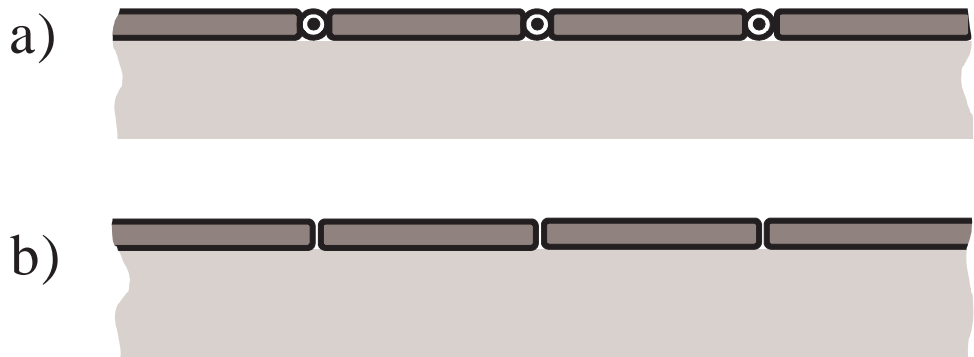}
\caption[]{The segmented elastic beam in contact with a 2D medium homogeneous in the longitudinal coordinate. The discontinuities allowing rotations under the bending moment (a) and entirely separated parts of the beam, which can interact with each other through the medium (b) $-$ as Professor Gennady Mishuris sees it. }
\label{fin}
\end{figure}

\vspace{35mm}

For the analytical study of such systems we use the approach earlier discussed in fracture mechanics where a crack was represented as the result of the action of generalized forces on the continuous medium, Slepyan (1981, \S 1.4; 2002, Sect. 5.7).

We use Lagrangian formulation considering the discontinuities as generalized strains created in an initially homogeneous structure by localized, self-equilibrated, generalized loads. Mathematically, this approach differs by using some higher order `deltas,' the derivatives of the  Dirac delta-function, higher than those related to real external localized forces.

For example, the functions $\Gd(x)$ and $\Gd'(x) = \D \Gd(x)/\D x$, corresponding to jump discontinuities in the transverse force and bending moment in the beam, make the internal force and moment discontinuous. These facts are well-known. At the same time, no one prevents us from continuing with the derivatives of higher orders taking generalized loads $\Gd''(x)$ and $\Gd'''(x)$. The latter corresponds to the structural discontinuities, the jumps in the inclination angle of the beam $w'(x)$ and its displacement $w(x)$, \res. Concerning the above considerations, we note that the delta-function can be defined as the derivative of a unite jump of discontinuity (see the related topics, e.g., in Bremermann, 1965).

Formally, the `loads' creating structural discontinuities do not affect the regular strain outside of the singular points but result in supersingular strain at these points. The latter, however, have no connection with the actual interface interaction, for which we introduce relations between the discontinuities and the corresponding force factors independently.

In doing so, we eliminate the obstacle for the use of the integral Fourier transform for the segmented structure with an adjoining medium, which reflects itself by Green's function. In the case of a uniform discrete distribution of the discontinuities, the coupled discrete and continuous transforms present in the analysis.

As a simple example consider the static problem for an elastic beam rested on Winkler's foundation.
\subsection{Elastic beam with a single hinge}
The Euler-Bernoulli equation for a uniform elastic beam on Winkler's foundation is
\beq D\f{\D^4 w(x)}{\D x^4} +\Gk w(x) =q(x)\,,~~~-\infty<x<\infty\,,\eeq{1}
where $w(x)$ is the displacement, $D$ is the bending stiffness, $\Gk$ is the bed stiffness and $q(x)$ is the external load.

Let there be an integrity violation at $x=0$, such that results in a jump discontinuity of the inclination angle (the displacement derivative). Denote it by $\Ga=w'(+0)-w'(x-0)$.

We assume that there exists a linear relation between the angle discontinuity and the bending moment
\beq \CM(\pm 0) = \CM(0) = - \Gk_M \Ga\,,~~~\Gk_M = \mbox{const} \ge 0\,.\eeq{2d}
In particular, the entire separation with respect to the rotation corresponds to $ \CM(\pm 0)= 0 \,(\Gk_M=0)$,  whereas$\Ga= 0 \,(\Gk_M=\infty)$ corresponds to the intact beam without the discontinuity.

Consider a function with the $\Ga$-discontinuity which satisfies the equation \eq{1} outside the singular point.
Substituting it in the equation \eq{1} we have the equality
\beq D\f{\D^4 w(x)}{\D x^4} +\Gk w(x) +D\Ga\Gd''(x) = q(x)\,.\eeq{E1}
This function represents the solution for the beam split into two parts as soon as we introduce the last term as a generalized load.

Recall that the known (linear) expression of the moment through the beam curvature
\beq \CM(x) =  - D\f{\D^2 w(x)}{\D x^2}\eeq{2da}
does not apply to the point $x=0$. The singularity at this point
\beq  \f{\D^2 w(x)}{\D x^2} = \Ga\Gd(x) ~~~(x=0)\eeq{2da1}
is only a consequence of the our way of the considerations.

The Fourier transform leads to
\beq w^F(k) = \f{q^F(k) - D\Ga k^2}{Dk^4 +\Gk}\,, \n \CM^F(k)=\f{D(q^F(k)k^2 -D\Ga k^4)}{Dk^4 +\Gk}+ D\Ga = \f{D( q^F(k)k^2 +\Gk\Ga) }{Dk^4 +\Gk}\,,\eeq{E2}
where in accordance with the above note, we excluded the singular term $ [- D\Ga\Gd(x)]^F(k) = - D\Ga$ from the regular expression for the bending moment. The relation \eq{2d} becomes
\beq \f{D}{2\pi}\inti \f{ q^F(k)k^2 +\Gk\Ga }{Dk^4 +\Gk}\D k = -\Gk_M\Ga\,.\eeq{E2a}
So, we find the value of the discontinuity as a function of the load and the hinge stiffness $\Gk_M$
\beq \Ga = - \f{1}{2\pi}\inti\f{q^F(k)k^2 \D k}{Dk^4 +\Gk}\gl(\f{\sqrt{2}}{4}\gl(\f{\Gk}{D}\gr)^{1/4} +\f{\Gk_M}{D}\gr)^{-1}\,.\eeq{E3}
The beam displacement $w(x)$, the Fourier transform of which is now fully defined in \eq{E2} with \eq{E3},
is plotted in \fig{aa3} for several values of $\Gk_M/D$  with $q(x)/D=\Gd(x)$ and $\Gk/D=1$.

\begin{figure}[h!] 

\vspace{-0mm}\hspace{-15mm}
\hspace{30mm}\includegraphics*[width=0.8\textwidth]{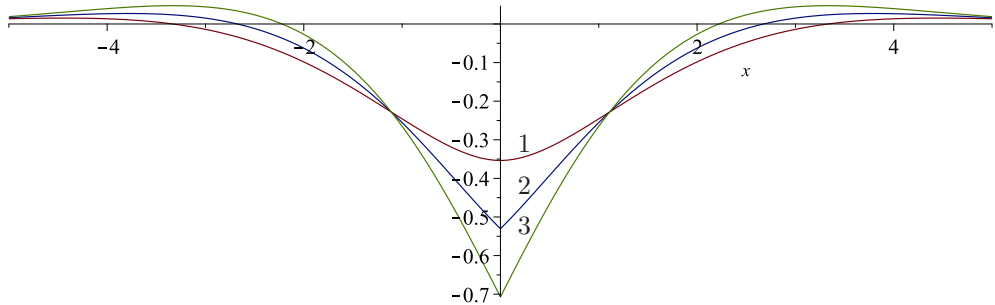}
\begin{picture}(0,0)(0,-100)
\put(-185,-39){{\footnotesize 1}}
\put(-185,-55){{\footnotesize 2}}
\put(-185,-70){{\footnotesize 3}}
\end{picture}
\vspace{0mm}
\caption[]{The beam displacement:  (1) the intact beam, $\Gk_M=\infty,  \Ga= 0$, (2) the partly separated, $\Gk_M/D = \sqrt{2}/4, \Ga = 1/2$, and (3)  the fully separated one, $\Gk_M=0, \Ga=1$.}
\label{aa3}
\end{figure}

Even in this simple example, the above way of the considerations appears preferable as compared to the usual analysis based on the examination of the beam parts separately. Indeed, in the latter case, the subsequent conjugation leads to four equations respective to the force, moment and displacement continuity and rotation discontinuity. In contrast, we have to determine only those discontinuities which do exist.  Namely, we have a single equation concerning the angle discontinuity, or two equations if both the angle and displacement have such. The continuity in other preserves itself automatically.

Below, after the general formulation, we consider some illustrative examples for the displacements and Floquet waves in the beam on Winkler's foundation, where the beam-medium interaction is local, for a Floquet gravity wave in a plate on deep water and Rayleigh-Floquet wave in an elastic plate on an elastic half-space. Note that in two last problems, the interactions with the media are nonlocal, and they can interact with each other through the medium even in the case where the beam parts entirely separated.

In the Appendix,  we present the dynamic elasticity wave equations developed for an elastic medium with discontinuities.

\section{General formulation}\label{gf}
So, we distinguish two formulations, physical, which we must eventually follow, and mathematical, which we use in solving the problems. In the former, we have the segmented structure, which parts under internal forces can have limited turn and shift relative to each other. In contrast, in the mathematical formulation, which is preferable for the analysis, we have the same but homogeneous structure, where the same as above discontinuities arise due to the action of self-equilibrated point-localized loads.

Since outside the singular points, the equations and boundary conditions adopted in these formulations coincide, the difference concerns only these points. In the physical formulation, we assume that there are relations between the structural discontinuities and related forces, whereas in the mathematical formulation we have supersingular strain there. Fortunately,
we can merely eliminate this difference. We ignore the singularity replacing it by the (physical) finite relations. Thereby, we make the mathematical solution entirely satisfying the physical formulation.

\subsection{Discontinuities and delta loads}
We consider a uniform linear differential equation of a general view including that for a 1D system coupled with a 2D homogeneous medium (homogeneous in the longitudinal direction)
\beq \sum_{n=1}^N a_nu^{(n)}(x) + \CG(x)\ast u(x) =q(x)\,,\eeq{gfe1}
where $a_n$ are constants, $C(x)$ is (inverse) Green's function as the response of the medium to $w(x)=\Gd(x)$, $\ast$ means the convolution, and $q(x)$ is a regular function. We assume that, under reasonable boundary conditions, this equation is satisfied by a regular function $u(x)$.  Note that
\beq \CG(x)\ast u(x) ~~\longrightarrow ~~ a_0u(x)~~~\mbox{if}~~~ \CG(x)=a_0\Gd(x)~~~(a_0 = \mbox{const})\,.\eeq{gfe2}
Also note that the equation can be one of a system. Therefore it can contain other functions. We, however, do not show it explicitly, since it does not affect the connections between the discontinuities and related deltas.

In dynamics, the functions $u$, $\CG$ and $q$ also depend on time, and $\ast$ means the convolution on both the coordinate and time. We will show this dependence explicitly in due time. In the current sections, it does not affect our considerations.

Now, let the $j$th order derivative have a discontinuity at $x=x_m$
\beq \GD u^{(j)}(x_m) = u^{(j)}(x_m+0)-u^{(j)}(x_m-0)\,.\eeq{gfe3}
We find
\beq  \sum_{n=1}^N a_nu^{(n)}(x)|_{x\ne x_m} + \CG(x)\ast u(x)\n + \GD u^{(j)}(x_m) \sum_{n=j+1}^N a_n\Gd^{(n-j-1)}(x-x_m)=q(x)\,.\eeq{gfe4}
Outside of the singular point $x=x_m$ this equation is satisfied by a regular function $u(x)$, but the equality does not hold at $x=x_m$ due to the presence of the singular term (the last sum). To compensate the singularity we introduce the same term as an additional discontinuity-associated load. The equation becomes
\beq \sum_{n=1}^N a_nu^{(n)}(x) + \CG(x)\ast u(x) = q(x)\n +\GD u^{(j)}(x_m) \sum_{n=j+1}^N a_n\Gd^{(n-j-1)}(x-x_m)\,.\eeq{gfe5}
The Fourier transform leads to the solution
\beq  u^F(k) = \gl(q^F(k) +\GD u^{(j)}(x_m) \sum_{n=j+1}^N a_n(-\I k)^{n-j-1}\gr)\E^{\I k x_m} \n
\times \gl(\sum_{n=1}^N a_n(-\I k)^n u^F(k) + \CG^F(k) \gr)^{-1}\,.\eeq{gfe6}

It is not difficult to see that this solution does correspond to the discontinuity. Up to a term, the inverse-transform integral of which converges uniformly (hence defining a continuous function)
\beq \gl(u^{(j)}\gr)^F = (-\I k)^j u^F(k) = \GD u^{(j)}(x_m)\f{1}{\pi}\int_0^\infty \f{\sin [k(x-x_m)]}{k}\D k\n = \f{1}{2}\GD u^{(j)}(x_m)\mbox{sign}(x-x_m)\,.\eeq{gfe7}

Note, that the above result corresponds to the prescribed discontinuities. However, in some cases, as shown below, the structural discontinuities break the continuity of the internal forces (as if a non-self-equilibrated action presents), which affects the analysis.

\subsection{Localised loads: forces and self-equilibrated actions}\label{lsel}
It may happen that a single discontinuity associates not only with a self-equilibrated action but a force or a moment. In this case,
another discontinuity should exist, such that the non-equilibrated effects cancel out. It follows from the equilibrium and the fact that only a self-equilibrated localized load produces the discontinuity.

Consider this issue by an example of the same elastic beam as above (see equation \eq{1}) but the beam under a tensile force, $T$. The equation for the continuous beam is
\beq D\f{\D^4 w(x)}{\D x^4} -T\f{\D^2 w(x)}{\D x^2}+\Gk w(x) =q(x)\,,~~~-\infty<x<\infty\,.\eeq{1tf}
Let both geometrical values $w(x)$ and $w'(x)$ have discontinuities at $x=0$. Under these (and only these) discontinuities the equation becomes
\beq D\f{\D^4 w(x)}{\D x^4} -T\f{\D^2 w(x)}{\D x^2}+\Gk w(x) =q(x)\n -D[\Ga\Gd''(x) +\Gb\Gd'''(x)]+T[\Ga\Gd(x) +\Gb\Gd'(x)]\,.\eeq{2atf}
Recall that $\Ga=\GD w'(0)$ and $\Gb = \GD w(0)$.

In this case, due to the presence of the tensile force, the localized external force $T\Ga$ and moment $T\Gb$ appear related to the discontinuity of $w'(0)$ and $w(0)$, \res. These forces, however, do not produce the discontinuities, because the latter are the results of self-equilibrated actions. The forces are the matter of equilibrium, and they must cancel out in the equation of equilibrium. It follows that the prescribed irregularities cause discontinuities in $w'''(0)$ and $w''(0)$, \res, such that
\beq \GD\CQ =- D\GD w'''(0) =  -T \Ga\,,~~~\GD\CM =- D\GD w''(0) =  -T \Gb\,.\eeq{4tf}
So, in the presence of the tensile force, the structure discontinuities cause the discontinuities in the bending moment and transverse forces in the beam.

Thus, the displacement (the Fourier transform of it) is
\beq w^F(k) = \f{q^F(k) - Dk^2 (\Ga-\I k \Gb)}{Dk^4 + Tk^2 +\Gk}\,.\eeq{2tf}

\subsection{The problem in the classical framework}
For the reader who would prefer traditional analysis another way of deriving the solution, \eq{gfe6}, is below presented, where we avoid consequent differentiation of the discontinuity and the necessity to use deltas at all.

Let a discrete set of the singular points, $x=x_m$, be associated with the equation \eq{gfe1}.
The Fourier transform of a single term over a single segment {\em between} the neighboring singular points, $x_m <x<x_{m+1}$, with the integration by parts is
\beq \int_{x_m+0}^{x_{m+1}-0}u^{(n)}(x)\E^{\I k x}\D x = (-\I k)^n u^{F_m}(k) +A_-+A_+\,,\n
A_- = - \sum_{\nu=0}^{n-1}(-\I k)^\nu u^{(n-1-\nu)}(x_{m}+0)\E^{\I k x_{m}}\,,\n A_+ = \sum_{\nu=0}^{n-1} (-\I k)^\nu u^{(n-1-\nu)}(x_{m+1}-0)\E^{\I k x_{m+1}}. \eeq{ftf3}
Note that the inverse transform results in a zero outside this segment.

The transform on the previous segment, $x_{m-1} <x< x_m$ gives us the same integrated terms at $x=x_m$ as $A_+$ but with $x=x_m$. Similarly, this action on the next segment, $x_{m+1} <x< x_{m+2}$ gives us at $x=x_{m+1}$ the same integrated terms as $A_-$  but with $x=x_{m+1}$. It follows that the integrated terms at $x=x_m$ and $x=x_{m+1}$  annihilate if the corresponding derivatives at these points are continuous. (Indeed, the continuous Fourier transform does not contain integrated terms.) If the left (right) integration limit goes to $\mp \infty$, it brings no integrated terms if the generalized Fourier transform is assumed (Bremermann, 1965).

Now, let there be the discontinuity, $\GD u^{(j)}(x_m)$. In this case, we obtain that its $m$-contribution to the right-hand side of the equation
\beq \GD u^{(j)}(x_m) a_n(-\I k)^{n-j-1}\E^{\I k x_m} \eeq{ftf3a}
coincide with that reflected in \eq{gfe6}. Making the transform for all $n$  and combining the finite-support transforms we obtain the solution \eq{gfe6}.

After these `classical' considerations we, however, should then discuss what happens at the discontinuity points (as we already did in the previous sections).

\subsection{Periodic array of the discontinuities}
In the case of a uniform distribution of a single discontinuity (as considered above in \eq{gfe3} - \eq{gfe6})
\beq \GD u^{(j)}(x) = \GD u^{(j)}(x_m)\,,~~~x_m=am\,,~~~a = \mbox{const}\,,\eeq{pard1}
we have the following right-hand part of the equation 
\beq S(x)=q(x)+\sum_{m=-\infty}^\infty \GD u^{(j)}(x_m)a_n(m)\Gd^{(n-j-1)}(x-x_m)\,,\eeq{pard2}
which in terms of the discrete Fourier transform becomes
\beq S^F(k) = q^F(k)+S^{Fd}(k)\,,\n
S^{Fd}(k) = \sum_{m=-\infty}^\infty \GD u^{(j)}(x_m)a_n(m) (-\I k)^{n-j-1}\E^{\I k x_m}\,.\eeq{pard3}
So, in this case, we have both the continuous and discrete Fourier transform coupled. With the aim to use the interface relations
like those in \eq{2dmm}, we have to rearrange the coupled transform to the discrete one.

\subsection{The Fourier transform: from continuous-discrete coupled\\ to discrete}\label{cdc}
Consider a function $f^F(k)$, which can be treated as the Fourier transform of a function $f(x)$
\beq f^F(k) = \inti f(x)\E^{\I k x}\D x\,.\eeq{ctd1}
Apparently, the corresponding discrete transform
\beq f^{Fd}(k) = \sum_{n=-\infty}^\infty f(an)\E^{\I k a n}\,,~~~a>0\,,\eeq{ctd2}
can be expressed from the continuous one in the following way
\beq f^{Fd}(k) = \f{1}{2\pi} \inti f^F(k') \sum_{n=-\infty}^\infty \E^{\I (k-k') a n}\D k' = \inti f^F(k') \GF(k-k')\D k'\,,\n
\GF(k-k')=\GF_+(k-k') +\GF_-(k-k')\,,\n \GF_+(k-k') =  \f{1}{2\pi} \f{1}{1-\exp[\I (k-k'+\I 0) a]}\,,\n \GF_-(k-k')= \f{1}{2\pi} \f{\exp[-\I (k-k'-\I 0) a]}{1-\exp[-\I (k-k'-\I 0) a]}\,,\n
\GF_+(k-k')+\GF_-(k-k') = \sum_{n=-\infty}^\infty \Gd[a(k-k') + 2\pi n]\,.\eeq{ctd3}
Thus, the expression is
\beq f^{Fd}(k)= \f{1}{a}\sum_{n=-\infty}^\infty f^F\gl(k+\f{2\pi n}{a}\gr)\eeq{ctd4}
It follows that the discrete transform defined by the last relation is a $2\pi/a$ periodic function of $k$, as well as the functions $\GF_\pm$, as they should be, and the latter represent functions regular in the upper/lower complex half-planes of $k$, \res. So, the functions $\GF_\pm$ are the kernels of the Cauchy-type integral (for the discrete transform)
\beq  f^{Fd}_\pm (k) = \inti f^F(k') \GF_\pm (k-k')\D k' = a\int_{-\pi/a}^{\pi/a}f^{Fd}(k')\GF_\pm(k-k')\D k'\,.\eeq{ctd5}

In the same manner, it can be seen that similar results correspond to a product of the continuous and discrete Fourier transforms. Due to the periodicity of the latter
\beq h^{Fd}(ak+2\pi n)=h^{Fd}(k)~~~\gl(f^F(k)=g^F(k)h^{Fd}(k)\gr)\,,\eeq{ctd6}
the corresponding discrete transform is
\beq f^{Fd}(k) =  \f{1}{a}h^{Fd}(k)\sum_{n=-\infty}^\infty g^F\gl(k+\f{2\pi n}{a}\gr)\,.\eeq{ctd7}

Concerning the summation, it should be noted that each part of the sum, $\sum_{n=-\infty}^{-1}$ and $\sum_{n=0}^\infty$, separately can diverge as well as the total sum as it is presented in \eq{ctd7}, while its another form, where the two sums are summed sequentially
\beq \sum_{n=-\infty}^\infty g^F\gl(k+\f{2\pi n}{a}\gr) = g^F(k)\n  + \sum_{n=1}^\infty \gl[g^F\gl(k+\f{2\pi n}{a}\gr) + g^F\gl(k-\f{2\pi n}{a}\gr)\gr]\n =\f{1}{2} \sum_{n=-\infty}^\infty \gl[g^F\gl(k+\f{2\pi n}{a}\gr) + g^F\gl(k-\f{2\pi n}{a}\gr)\gr]\,,\eeq{sums1}
converges. This can happen for an asymptotically odd function of $n$ where
\beq g^F\gl(k-\f{2\pi n}{a}\gr) \sim - g^F\gl(k+\f{2\pi n}{a}\gr) ~~~(n\to\infty)\,.\eeq{sums2}
Note that if the sum $\sum_{n=-\infty}^\infty g^F(k+2\pi/a)$ converges, then
\beq \sum_{n=-\infty}^\infty g^F\gl(k+\f{2\pi n}{a}\gr) = \sum_{n=-\infty}^\infty g^F\gl(k-\f{2\pi n}{a}\gr)\,,\eeq{sums3}
and the representation \eq{sums1} is identical to that in \eq{ctd7}. Otherwise, only the rule \eq{sums1} should be accepted, and we use it for $Q_q^F(k)$ in the beam related section.

Note that due to the presence of $(k+2\pi m/a)^4$ term, the infinite sums in  \eq{ctd7}, at least in the form in \eq{sums1}, converge fast.

\section{The beam-plate problems}\label{bpp}
We now consider the beam with two types of the discontinuity, the jumps of both the displacement derivative and the displacement itself
\beq \Ga_n = w'(x_n+0)-w'(x_n-0)\,,~~~\Gb_n = w(x_n+0)-w(x_n-0)\,,\n n= 1, 2, ..., N\,.\eeq{gf1a}
Here we do not specify the medium with which the beam is coupled. In the same way as in the above example, we get the equation
\beq D\f{\D^4 w(x)}{\D x^4} +\CG(x)\ast w(x) =q(x)+D\sum_{n}\Ga_n\Gd''(x-x_n)+\Gb_n\Gd'''(x-x_n),\eeq{gf2a}
where $\CG(x)$ is (inverse) Green's function as the response to the displacement $w(x)=\Gd(x)$, and $\ast$ means the convolution. In dynamics, the discontinuity parameters $\Ga$ and $\Gb$ become functions of time as well as $w = w(x,t)$, $\CG = \CG(x,t)$, and the convolution also includes the beam inertia. However, in this section, we do not show the time-dependence explicitly, since it does not affect our reasoning.

The Fourier transform results in
\beq [Dk^4 +\CG^F(k)]w^F(k) =q^F(k)-D k^2\sum_{m}(\Ga_m-\I k\Gb_m)\E^{\I k x_m}\,.\eeq{gf2}

The dependencies for the displacement and its derivatives for the singular interface located at $x=0$  are as follows (the relations for other points differ only by the exponential multipliers as in \eq{gf2})
\beq  w^F(k) = \f{q^F(k) -  D k^2(\Ga -\I k  \Gb) }{ Dk^4 +\CG^F(k)}\,,\n
(w')^F(k)  =\underline{\Gb}  - \f{\I k q^F(k) +\Gb \CG^F(k) -D\Ga\I k^3}{Dk^4 +\CG^F(k)}\,,\n
(w'')^F(k) = \underline{\Ga-\I k \Gb} + (w''_{reg})^F(k)\,,   \n (w''')^F(k)= \underline{-\I k\Ga-k^2 \Gb} + (w'''_{reg})^F(k)\,,\eeq{2a1}
with
\beq (w''_{reg})^F(k)=  - \f{k^2 q^F(k)+(\Ga-\I k\Gb)\CG^F(k)}{Dk^4+\CG^F(k)}\,,\n
 (w'''_{reg})^F(k)=  \f{\I k^3 q^F(k)+(\I k\Ga+ k^2\Gb)\CG^F(k)}{Dk^4 +\CG^F(k)}\,,\eeq{2a}
where the underlined terms correspond to the deltas at $x=0$, which do not affect the forces and should be ignored as discussed in \az{gf}.

Expressions for the total bending moment and the transverse force due to the external force and the discontinuities at $x=x_m,\, m=1, ..., N,$ follow  from these relations as
\beq \CM^F(k) =M^F_q(k) + \sum_{m=1}^N\gl[ \Lambda^F_\Ga(k)\Ga_m + \Lambda^F_{\Ga\Gb}(k)\Gb_m\gr]\E^{\I x_m}\,,\n
\CQ^F(k) =Q^F_q(k) + \sum_{m=1}^N \gl[ \Lambda^F_{\Ga\Gb}(k)\Ga_m +\Lambda^F_\Gb(k) \Gb_m\gr]\E^{\I x_m}\,,\eeq{gf3}
with
\beq M_q^F(k) = \f{D k^2 q^F(k)}{Dk^4 +\CG^F(k)}\,,~~~ Q_q^F(k) = - \f{D\I k^3 q^F(k)}{Dk^4 +\CG^F(k)}\,,\n
\Lambda^F_\Ga(k) = \f{D\CG^F(k)}{Dk^4 +\CG^F(k)}\,,~~~\Lambda^F_{\Gb}(k)= - \f{D k^2\CG^F(k)}{Dk^4 +\CG^F(k)}\,, \n
\Lambda^F_{\Ga\Gb}(k) = - \f{D\I k\CG^F(k)}{Dk^4 +\CG^F(k)}\,.\eeq{gf3a}

The relations between the discontinuities $\Ga$ and $\Gb$ on the one hand, and the corresponding forces, $\CM$ and $\CQ$ on the other hand are
\beq   \CM(x_n)  = - \Gk_{Mn} \Ga_n\,,~~~  \CQ(x_n)  =  \Gk_{Qn} \Ga_n\,.\eeq{2dmm}

Thus, we have $2N$ interface conditions \eq{2dmm}, which allow us to determine the discontinuity parameters $\Ga_n$ and $\Gb_n$. After this, the expression in \eq{gf2} uniquely defines the beam displacement.

\subsection{Uniformly distributed discontinuities}
Here, we reduce the formulation to the case of uniformly distributed  discontinuities:  $x_n = an, n=0, \pm 1, ..., a = $ const $>0$ with $\Gk_{(M,Q)n}=\Gk_{(M,Q)}$. In this case, the sums on $m$ in \eq{gf2} and \eq{gf3} become the Fourier discrete transforms
\beq [Dk^4 +\CG^F(k)]w^F(k) =q^F(k)-D k^2\!\gl(\Ga^{Fd}(k) -\I k\Gb^{Fd}(k)\gr)\,,\eeq{gf2d}
where for a function $f(n)$
\beq f^{Fd}(k) = \sum_{n=-\infty}^\infty f(n)\E^{\I a n}\,.\eeq{dFt}
The expressions for the bending moment and transverse force \eq{gf3} become
\beq \CM^F(k) =M^F_q(k) +  \Lambda^F_\Ga(k)\Ga^{Fd}(k) + \Lambda^F_{\Ga\Gb}(k)\Gb^{Fd}(k)\,,\n
\CQ^F(k) =Q^F_q(k) +  \Lambda^F_{\Ga\Gb}(k)\Ga^{Fd}(k) +\Lambda^F_\Gb(k) \Gb^{Fd}(k)\,,\eeq{gf3d}
where the coefficients $\Lambda^F(k)$ are defined in \eq{gf3a}.
Also, as follows from \eq{2dmm}
\beq  \CM^{Fd}(k)  = - \Gk_{M} \Ga^{Fd}(k)\,,~~~ \CQ^{Fd}(k)  =  \Gk_{Q} \Ga^{Fd}(k)\,.\eeq{2dmmp}

Now, referring to \eq{2dmmp} and \eq{ctd4}, we get the following equations concerning the discontinuity parameters $\Ga$ and $\Gb$
\beq \CM^{Fd}(k) =M^{Fd}_q(k) +  \Lambda^{Fd}_\Ga(k)\Ga^{Fd}(k) + \Lambda^{Fd}_{\Ga\Gb}(k)\Gb^{Fd}(k) = - \Gk_M \Ga^{Fd}(k),\n
\!\! \CQ^{Fd}(k) =Q^{Fd}_q(k) +  \Lambda^{Fd}_{\Ga\Gb}(k)\Ga^{Fd}(k) +\Lambda^{Fd}_\Gb(k) \Gb^{Fd}(k) = \Gk_Q \Gb^{Fd}(k),\eeq{gf3dd}
where
\beq \Lambda^{Fd}_{\Ga,\Ga\Gb,\Gb}(k)=\f{1}{a}\sum_{n=-\infty}^\infty \Lambda^{F}_{\Ga,\Ga\Gb,\Gb}\gl(k+\f{2\pi n}{a}\gr)\,.\eeq{gf3dda}

As the discontinuity parameters are determined from these two equations, we can go back to the beam displacement defined in \eq{gf2d}. However, the homogeneous solution (the dispersion relation in the case of waves) follows from  \eq{gf3dd} directly. If $q=0$ a nontrivial solution satisfies the relation as
\beq (\Lambda^{Fd}_\Ga(k)+\Gk_M)(\Lambda^{Fd}_\Gb(k)-\Gk_Q)-(\Lambda^{Fd}_{\Ga\Gb}(k))^2=0\,,\eeq{gos}
and in the case where $\Gb^{Fd} = 0~(\Gk_Q=\infty)$, it reduces to
\beq \Lambda^{Fd}_\Ga(k)=-\Gk_M\,.\eeq{gos1}

At the limit $\Gk_M=\infty, \Gk_Q=\infty$, we return to the continuous beam with the Floquet-type  dispersion relation
\beq \prod_{n=-\infty}^\infty \gl[D\gl( k +\f{2\pi n}{a}\gr)^4 + \CG^F\gl( k +\f{2\pi n}{a}\gr)\gr] =0\,.\eeq{fwdrib}
This expression represents the set of modes indistinguishable at the discrete set of points $x= an$, where the Floquet wave is defined. So at this limit, the set is identical to the single dispersion relation
\beq Dk^4 + \CG^F(k) =0\,,~~~-\pi < ak \le \pi\,.\eeq{drib}

Summarising, the general procedure is as follows. Assuming Green's function $\CG^F(k)$ is given, first the $\Lambda^{Fd}$-functions should be determined based on \eq{gf3a} and \eq{ctd4}
\beq  \Lambda^{Fd}_\Ga(k)=\f{1}{a} \sum_{n=-\infty}^\infty \f{D\CG^F(k')}{D(k')^4 +\CG^F(k')}\,,\n
\Lambda^{Fd}_{\Gb}(k)= -\f{1}{a}\sum_{n=-\infty}^\infty  \f{D (k')^2\CG^F(k')}{D(k')^4 +\CG^F(k')}\,,\n
\Lambda^{Fd}_{\Ga\Gb}(k)= -\f{1}{a}\sum_{n=-\infty}^\infty  \f{D\I k'\CG^F(k')}{D(k')^4 +\CG^F(k')}\,,
~~~k' = k +\f{2\pi n}{a}\,.\eeq{gpr1}

Then, if the complete solution is required, similar transformation is to be applied for the load functions in \eq{gf3a}
\beq M^{F}_q(k)~~ \longrightarrow~~M^{Fd}_q(k)\,,~~~Q^{F}_q(k)~~ \longrightarrow~~Q^{Fd}_q(k)\eeq{gpr2}
based on \eq{ctd4} and \eq{sums1}, \res.

As a result, for the case $q=0$  we obtain the `dispersive relation' \eq{gos} defining the Floquet wavenumber $k$. Otherwise, the discontinuity parameters $\Ga^{Fd}(k)$ and $\Gb^{Fd}(k)$ can be found from \eq{gf3dd}.

Finally, the beam displacement and internal forces are defined in the form of (continuous) Fourier transform in \eq{gf2d} and \eq{gf3d}.

Below we present some illustrative examples assuming that the beam (or plate) has only angle discontinuities periodically distributed, whereas the displacement is continuous. In this case, the system \eq{gf3dd} reduces to a single equation as
\beq M^{Fd}_q(k) +  (\Lambda^{Fd}_\Ga(k)+\Gk_M)\Ga^{Fd}(k)  =0\,.\eeq{smc}

\section{The periodically segmented beam on elastic foundation}
\subsection{General  dependencies}
In this case, $\CG^F(k)=\Gk = $ const. We assume that the beam is under the force $q=P\Gd(x), q^F(k)=P$. The functions in  \eq{gf3} become
\beq M_q^{Fd}(k) =  aPS_2\,,~~~\Lambda^{Fd}_\Ga(k) =  \f{D S_0}{a}\,,~~~\Lambda^{Fd}_{\Ga\Gb}(k) = - \f{\I D S_1}{a^2}\,,\n
Q_q^{Fd}(k) = -\I P S_3\,,~~~\Lambda^{Fd}_\Gb(k)= - \f{D S_2}{a^3}\,,\eeq{sum1}
with
\beq S_0= \sum_{n=-\infty}^\infty\f{\Gk}{D(k+2\pi n/a)^4 +\Gk}=
\f{1}{2}\Re (\hat{\Gk}^{1/4}\Psi)\,,\n
S_1=\sum_{n=-\infty}^\infty \f{a(k+2\pi n/a) \Gk}{D(k+2\pi n/a)^4 +\Gk}\n = \f{\sqrt{\hat{\Gk}}}{2}\f{\sinh(\tau)\sin(\tau)\sin(ak)}{(\cosh(\tau)\cos(\tau)-\cos(ak))^2+(\sinh(\tau)\sin(\tau))^2}\,,\n
S_2=\sum_{n=-\infty}^\infty\f{(k+2\pi n/a)^2\Gk}{a^2(D(k+2\pi n/a)^4 +\Gk)}= \f{1}{2}\Im(\hat{\Gk}^{-1/4}\Psi)\,,\n
S_3=\sum_{n=-\infty}^\infty  \f{1}{2a}\gl(\f{ (k+2\pi n/a)^3D }{D(k+2\pi n/a)^4+\Gk}-\f{ (-k+2\pi n/a)^3D}{D(-k+2\pi n/a)^4 +\Gk}\gr)\n
 = \f{1}{2}\Re \f{\sin(ak)}{\cosh(\sqrt{\I}\hat{\Gk}^{1/4})-\cos(ak)}\,,\n
\Psi = \f{\sqrt{\I}\,\sinh(\sqrt{\I}\,\hat{\Gk}^{1/4})}{\cosh(\sqrt{\I}\,\hat{\Gk}^{1/4})-\cos(a k)}\,,~~~\hat{\Gk}=\f{\Gk a^4}{D}\,,~~~
\tau = \f{\sqrt{2}}{2}\hat{\Gk}^{1/4}\,.\eeq{sum2}

\subsection{The beam in statics}
The first equation in \eq{gf3dd} becomes
\beq aPS_2 +\gl(\f{D S_0}{a} +\Gk_M\gr)\Ga^{Fd}=0\,,~~~\Longrightarrow~~~\Ga^{Fd}=-\f{\hat{P}S_2}{ S_0+\hat{\Gk}_M}\,,\eeq{es4}
and
\beq w^F(k) =  \f{P-Dk^2\Ga^{Fd}}{Dk^4 + \Gk}
=\f{\hat{P}a^2}{(ak)^4 +\hat{\Gk}}\gl(1+\f{(ak)^2S_2}{S_0+\hat{\Gk}_M}\gr)\eeq{es5}
with
\beq \hat{P}= \f{a^2P}{D}\,,~~~\hat{\Gk}= \f{a^4\Gk}{D}\,,~~~\hat{\Gk}_M= \f{a\Gk_M}{D}\,.\eeq{es6}

The beam displacement corresponding to $\hat{P}=1$ and some values of $\hat{\Gk}_M$ for $\hat{\Gk}= 1$ and $\hat{\Gk}= 100$ is presented in \fig{2} and \fig{3}, \res.

\begin{figure}[h!] 

\vspace{0mm}\hspace{-15mm}
\hspace{30mm}\includegraphics*[width=0.8\textwidth]{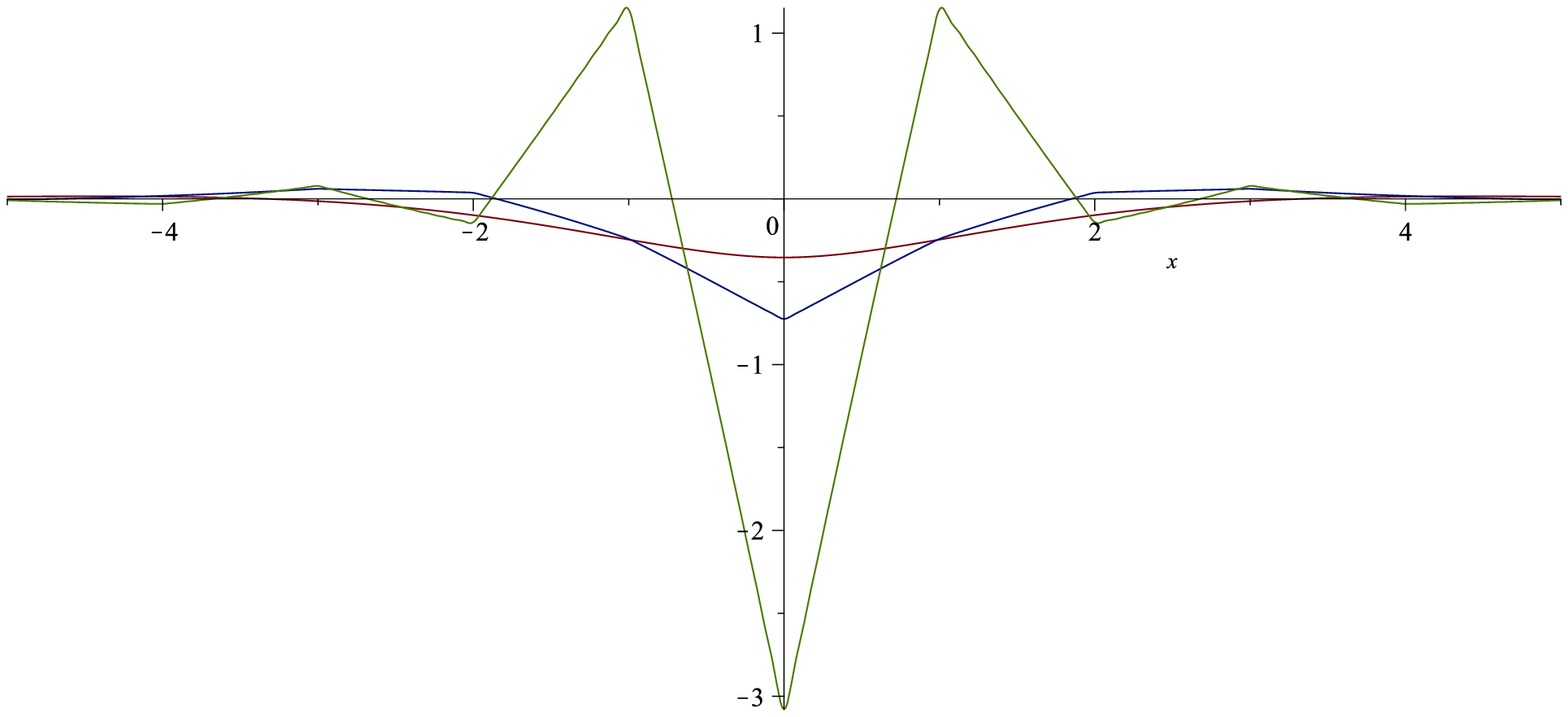}
\begin{picture}(0,0)(0,-100)
\put(-185,13){{\footnotesize 1}}
\put(-185,-8){{\footnotesize 2}}
\put(-185,-41){{\footnotesize 3}}
\end{picture}
\vspace{0mm}
\caption[]{The beam displacement for $\hat{P}=\hat{\Gk}=1$ and different values of the hinge stiffness:  $\Gk_M=\infty$ (1), $\Gk_M = 1/2$ (2) and $\Gk_M=0$ (3).}
\label{2}
\end{figure}

\begin{figure}[h!] 

\vspace{-0mm}\hspace{-15mm}
\hspace{30mm}\includegraphics*[width=0.8\textwidth]{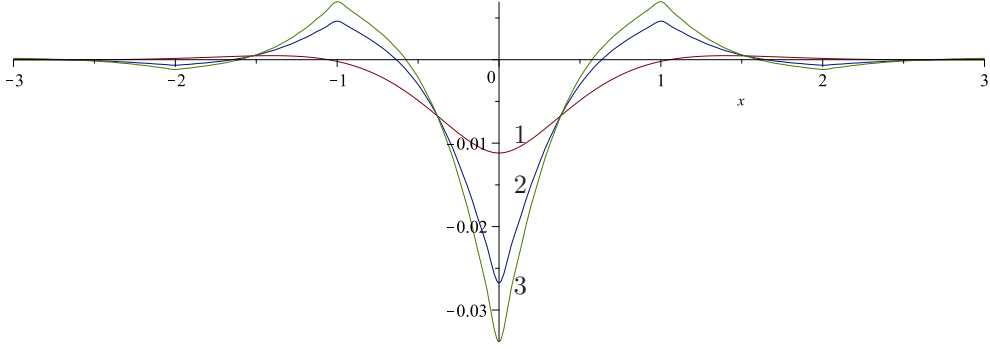}
\begin{picture}(0,0)(0,-100)
\put(-185,-23){{\footnotesize 1}}
\put(-185,-42){{\footnotesize 2}}
\put(-185,-80){{\footnotesize 3}}
\end{picture}
\vspace{0mm}
\caption[]{The beam displacement for $\hat{P}=\hat{\Gk}=100$: and different values of the hinge stiffness:  $\Gk_M=\infty$ (1), $\Gk_M = 1/2$ (2) and $\Gk_M=0$ (3).}
\label{3}
\end{figure}

\subsection{The beam in oscillations}
In a general dynamic case, we replace
 \beq \CG^F(k) ~~ \longrightarrow ~~ \Gr S s^2 + \CG^{LF}(s,k)\,,\eeq{6}
 where $\Gr$, $S$ and $s$ a are the beam material density and the cross-section area of the beam and the parameter of the Laplace transform on time $t$, \res, and $\CG(t,x)$ corresponds to the medium dynamic response to $w(t,x) = \Gd(t)\Gd(x)$.

 For the case of sinusoidal oscillations where the functions contain multiplier $\exp(\I\Go t)$, omitting the latter we return to the previous formulation with the change
\beq \Gk ~~ \longrightarrow ~~ \Gr S(0+\I\Go)^2 +\Gk \,,~~~ \hat{\Gk}~\longrightarrow ~
(0+\I\hat{\Go})^2 + \hat{\Gk}\,,~~~\hat{\Go}=\f{ a^2\Go}{r c}\,,\eeq{o1}
where $r$ is the radius of inertia of the beam cross-section area and $c=\sqrt{E/\Gr}$ is the longitudinal wave speed in the corresponding elastic rod. The homogeneous relation \eq{gos1} yields
\beq ak = \arccos\gl(\cos \Gt - \f{\Gt \sin\Gt}{2\hat{\Gk}_M}\gr)\,,~~~\Gt=(\hat{\Go}^2-\hat{\Gk})^{1/4}\,.\eeq{o2}
The dependence $ak(\hat{\Go})$ for $\hat{\Gk}=1$ and several values of the hinge stiffness $\hat{\Gk}_M$ is plotted in \fig{wo10}, \fig{wo1} and \fig{wo01}, where the dependence for the intact beam, $ak = (\hat{\Go}^2-\hat{\Gk})^{1/4}$ ($\hat{\Gk}_M=\infty$) is shown too.

\begin{figure}[h!]

\vspace{-0mm}\hspace{-15mm} 
\hspace{30mm}\includegraphics*[width=0.8\textwidth]{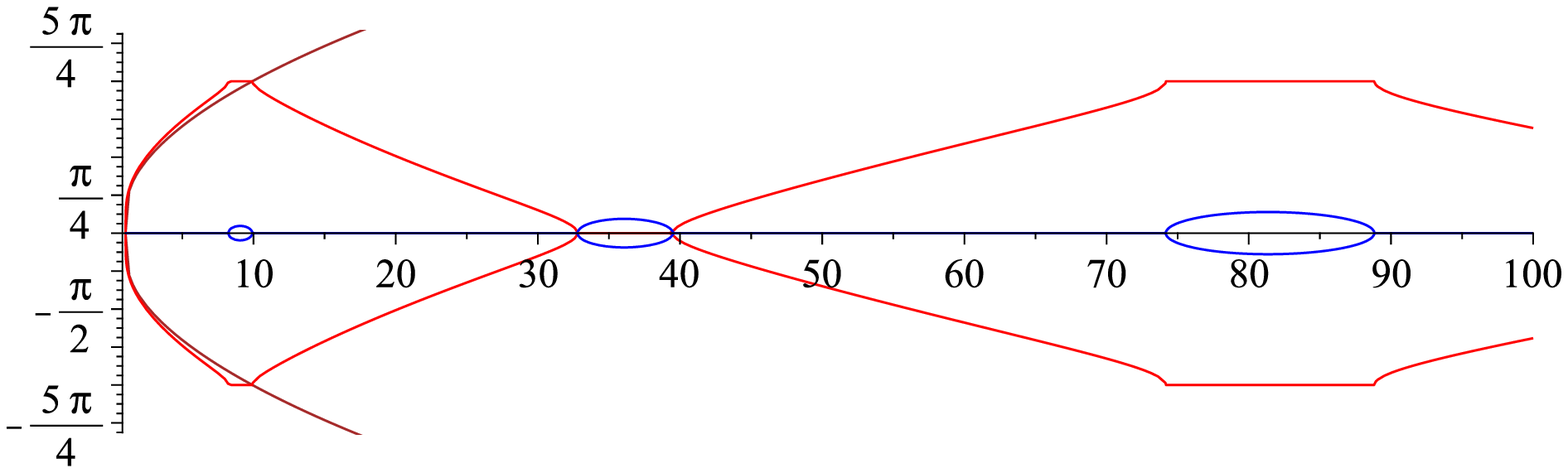}
\begin{picture}(0,0)(0,-100)
\put(-230,-34){{\footnotesize 1}}
\put(-284,-20){{\footnotesize 2}}
\put(-294,-85){{\footnotesize 3}}
\put(-364,-20){{$ak$}}
\put(-135,-56){$\hat{\Go}$}
\end{picture}
\vspace{0mm}
\caption[]{The beam on the elastic foundation. The Floquet normalized wavenumber $ak$ for $\hat{\Gk}_M=10$ and $\hat{\Gk}=1$:  (1) $a \Re k$ (red),~ (2) $a \Im k$ (blue) and (3) the regular normalized wavenumber for the intact beam, $\hat{\Gk}_M=\infty$ (brown)}.
\label{wo10}
\end{figure}

\begin{figure}[h!]

\vspace{-0mm}\hspace{-15mm} 
\hspace{30mm}\includegraphics*[width=0.8\textwidth]{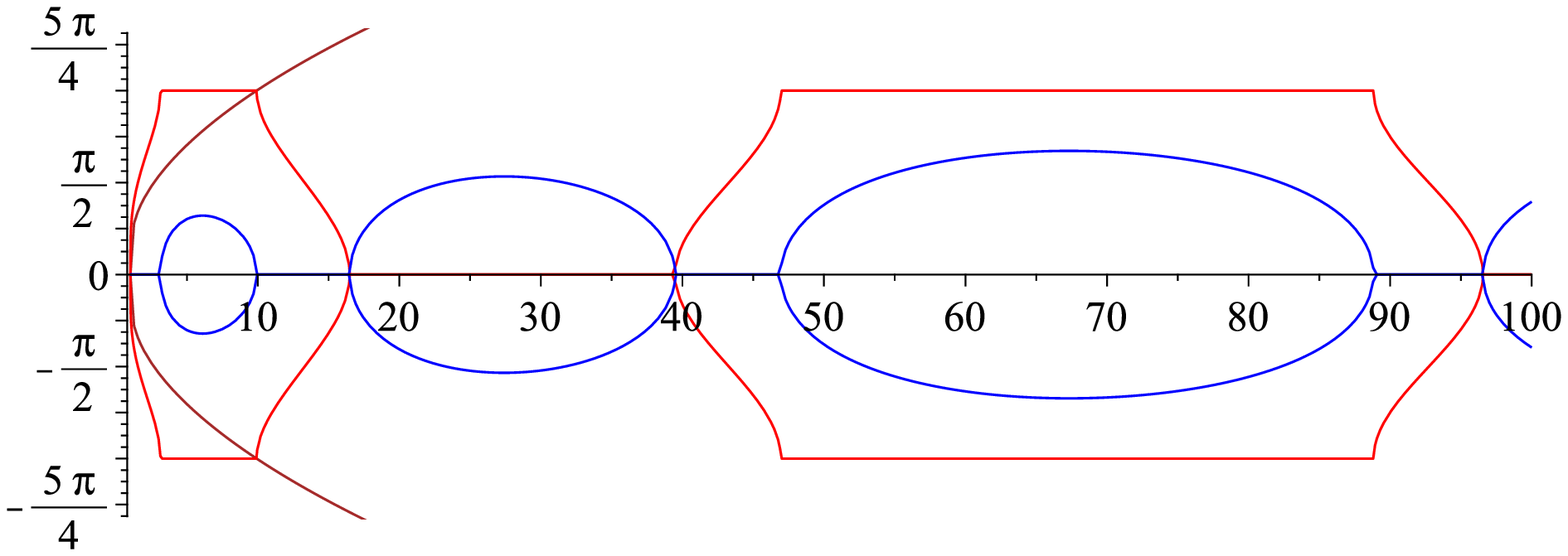}
\begin{picture}(0,0)(0,-100)
\put(-315,-64){{\footnotesize 1}}
\put(-260,-6){{\footnotesize 2}}
\put(-294,-85){{\footnotesize 3}}
\put(-364,2){{$ak$}}
\put(-133,-50){$\hat{\Go}$}
\end{picture}
\vspace{0mm}
\caption[]{The beam on the elastic foundation. The Floquet normalized wavenumber $ak$ for $\hat{\Gk}_M=1$ and $\hat{\Gk}=1$:  (1) $a \Re k$ (red),~ (2) $a \Im k$ (blue) and (3) the regular normalized wavenumber for the intact beam, $\hat{\Gk}_M=\infty$ (brown).}
\label{wo1}
\end{figure}

\begin{figure}[h!]

\vspace{-0mm}\hspace{-15mm}
\hspace{30mm}\includegraphics*[width=0.8\textwidth]{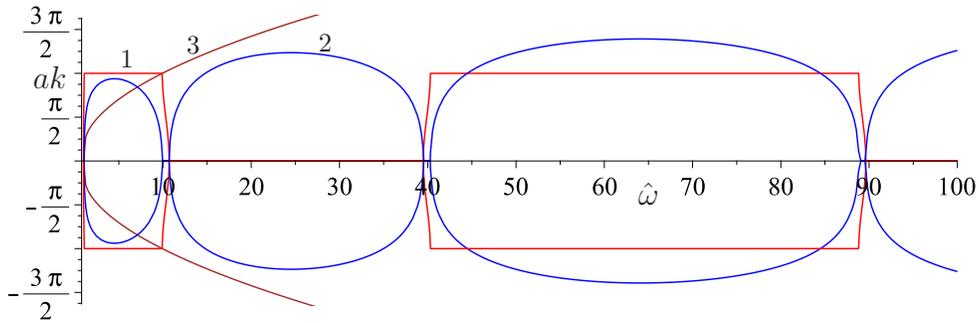}
\begin{picture}(0,0)(0,-100)
\put(-330,0){{\footnotesize 1}}
\put(-255,6){{\footnotesize 2}}
\put(-305,5){{\footnotesize 3}}
\put(-363,-8){{$ak$}}
\put(-134,-52){$\hat{\Go}$}
\end{picture}
\vspace{0mm}
\caption[]{The beam on the elastic foundation. The Floquet normalized wavenumber $ak$ for $\hat{\Gk}_M=0.1$ and $\hat{\Gk}=1$:  (1) $a \Re k$ (red),~ (2) $a \Im k$ (blue) and (3) the regular normalized wavenumber for the intact beam, $\hat{\Gk}_M=\infty$ (brown).}
\label{wo01}
\end{figure}

Can be seen that the band gaps grow as $\hat{\Gk}_M$ decreases and as $\hat{\Gk}_M\to 0$ cover almost whole domain, except narrowing vicinities of
\beq \hat{\Go} = \sqrt{(m\pi)^4+1}\,,~~~m=1, 2, ...\,.\eeq{o3}

\vspace*{20mm}

\section{A periodically segmented elastic plate on deep water}
\subsection{Gravity waves}
We consider a floating elastic plate rested on deep incompressible water, $z<0$. The formulation for the continuous plate coincides with that for capillary waves (ripples), see, e.g., Lighthill (1978), if the pressure by the surface tension term is replaced by the one corresponding to the plate: $Tk^2\longrightarrow Dk^4$. So, we have
\beq \CG^F(k) =\Gr_0  \gl(g - \f{\Go^2}{\sqrt{0+k^2}}\gr)\,,\eeq{7}
where $\Gg_0$ and $g$ are the water density and the acceleration due to gravity, \res. The frequency in the case of the intact plate is defined as
\beq \Go^2 = \gl(g + \f{D k^4}{\Gr_0}\gr)\sqrt{0+k^2}\,.\eeq{7aa}

Consider the homogeneous problem ($q=0$).
The function $\Lambda^{Fd}_\Ga(k)$ \eq{gpr1} follows  as
\beq \Lambda^{Fd}_\Ga(k)=\f{D}{a} \sum_{n=-\infty}^\infty \f{\hat{\CG}^F(\zeta)}{\zeta^4 +\hat{\CG}^F(\zeta)}\,,~~~
\hat{\CG}^F(\zeta) =C_g -\f{12h\Gr_0}{a\Gr \hat{\Gg}}\hat{\Go}^2\,,\n
C_g=\f{a^4\Gr_0g}{D}\,,~~~\hat{\Gg}=\sqrt{0+\zeta^2}~~~\hat{\Go}=\f{a^3\Go}{h^2c}\,,~~~\zeta=ak +2\pi n\,.\eeq{8}
We come to the dispersion relation  \eq{gos1}. In the non-dimensional form, it is
\beq \sum_{n=-\infty}^\infty \f{\hat{\CG}^F(\zeta)}{\zeta^4 +\hat{\CG}^F(\zeta)}) = - \hat{\Gk}_M\,,~~~\hat{\Gk}_M=\f{a \Gk_M}{D}\,,\eeq{9}
The relation for the intact plate, $\hat{\Gk}_M=\infty$, follows as
\beq \hat{\Go} = \sqrt{\f{\Gr a}{12\Gr_0h}\gl((ak)^4 +C_g\gr)\sqrt{0+(ak)^2}}\,.\eeq{ipdd}

The dispersion dependencies for several values of the hinge stiffness are shown in \fig{wg-inf}. It can be seen how the frequency maximal value decreases from $\hat{\Go}_{max} \approx 23$ for the intact plate ($\Gk_M=\infty$) to $\hat{\Go}_{max} \approx 4$ for the case of the perfect hinges, $\hat{\Gk}_M=0$. The plots correspond to $C_g= 3$ and $12 h\Gr_0 =0.6 a\Gr $. Also, the discontinuities entail the band gap extending as $\Gk_M$ decreases. Its upper bound is
\beq \Go^*_{max} = \sqrt{\f{\pi g}{a} + \f{\pi^5 D}{\Gr_0 a^5}}\eeq{ubbg}
(it is equal to $\Go_{max}$ at $\Gk_M=\infty$), whereas $\Go^*_{min}$ is the frequency at $k=\pi/a$ corresponding to the stiffness $\Gk_M$ under consideration. Graphs of the bounds are plotted in \fig{w-band}. Note that $\Go^*_{max}$  independency of $\Gl_M$  follows from the fact that for $k=\pi/a$  the terms $n=0 $ and $n=-1$ (in the sum by $n$) coincide.

\begin{figure}[h!]
\vspace{-0mm}\hspace{-15mm}
\hspace{30mm}\includegraphics*[width=0.4\textwidth]{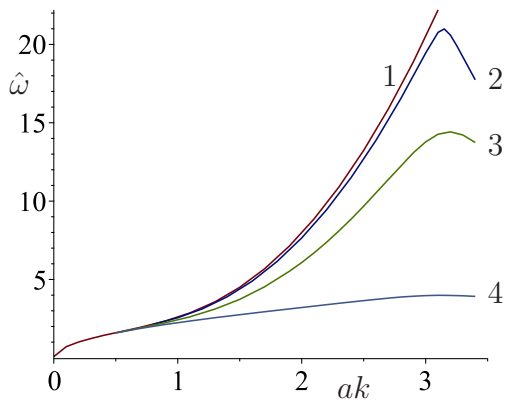}
\begin{picture}(0,0)(0,-100)
\put(-188,17){{$\hat{\Go}$}}
\put(-64,-98){{$ak$}}
\put(-47,20){{$1$}}
\put(-7,19){{$2$}}
\put(-7,-6){{$3$}}
\put(-7,-62){{$4$}}
\end{picture}
\vspace{0mm}
\caption[]{The $2\pi$-periodic Floquet dispersive dependence $\hat{\Go}(-ak)=\hat{\Go}(ak)$ for the gravity wave in a fluid covered by an intact elastic plate. The regular dependence corresponding to the intact plate, $\Gk_Q=\Gk_M =\infty$, coincides with the plotted one (1) and continues for greater $k$ in accordance with \eq{ipdd}. The dependencies for finite values of $\Gk_M =\infty$ ($\Gk_Q=\infty$) plotted for $\Gk_M =10$ (2), $\Gk_M =1$ (3) and $\Gk_M =0$ (4)}.
\label{wg-inf}
\end{figure}

\begin{figure}[h!]

\vspace{10mm}\hspace{-15mm}
\hspace{30mm}\includegraphics*[width=0.4\textwidth]{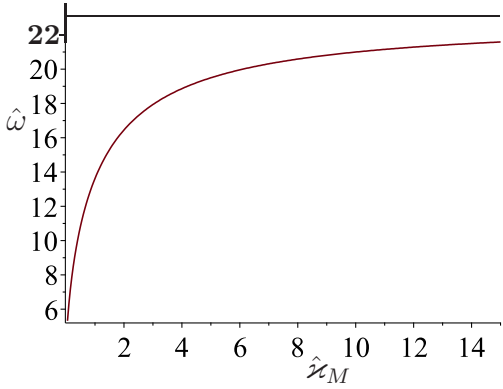}
\begin{picture}(0,0)(0,-100)
\put(-171.5,44){\line(1,0){164}}
\put(-171.7,34){\line(0,1){15}}
\put(-81,-93){{$\hat{\Gk}_M$}}
\put(-194,01){$\hat{\Go}$}
\put(-190,33.8){\fontsize{11.2}{1} {\bf 22}}
\put(-174,36.8){\line(1,0){2}}
\end{picture}
\vspace{0mm}
\caption[]{The $2\pi$-periodic Floquet dispersive dependence $\hat{\Go}(-ak)=\hat{\Go}(ak)$ for the gravity wave in a fluid covered by an elastic plate with the periodically placed discontinuities. The band gap upper and lower bounds.}
\label{w-band}
\end{figure}

\vspace{10mm}
\section{The plate in sliding contact with an elastic medium}
The waves in the homogeneous flexible plate contacting with the elastic half-space where studied in Achenbach and Epstein  (1967) and Ozisik and Akbarov (2003). Below we consider this problem for the segmented plate as in the previous section, but for an elastic half-plane with the gravity neglected.

Green's function for the corresponding plane problem can be found, e.g., in Slepyan (2002, Sec. 9.1.3). For our case where $w=u_2=\Gd(x)\Gd(t), \,w^{FL}(s,k) =1$
\beq \CG_0^{LF}(s,k)= \Gs_{22}^{LF} = -\f{1}{S^{LF}_{22}} = \f{\Gm c_2^2 R}{s^2n_1}~~~(\Gs_{12}=0)\,,\n
R = (n_2^2+k^2)^2-4k^2n_1n_2\,,\n n_{1,2}=\sqrt{k^2 + \f{s^2}{c_{1,2}^2}}\,,~~~c_1^2=\f{\Gl +2\Gm}{\Gr_1}\,,~~~c_2^2=
\f{\Gm}{\Gr_1}\,,\eeq{em1}
where $\Gl$ and $\Gm$ are the Lam\'{e} elastic constants and $\Gr_1$ is the medium density.

 For the complex wave with dependence on time as $\E^{\I\Go t}$, the above relations remain valid as $s \longrightarrow 0+\I \Go$. We come to the same expression for $ \Lambda^{Fd}_\Ga(k)$ as in \eq{8}, but with the following function $\CG^F(\zeta)$
\beq \hat{\CG}^F(\zeta) = - A\hat{\Go}^2 - \f{B[(2\zeta^2 - \hat{\Go}^2)^2 -4\zeta^2\sqrt{\zeta^2 - \hat{\Go}^2/4}\sqrt{\zeta^2 - \hat{\Go}^2/4}]}{\hat{\Go}^2\sqrt{\zeta^2 - \hat{\Go}^2/4}}\,,\n
A=\f{12c_2^2a^2}{c^2h^2}\,,~~~B=\f{12\Gm(1-\nu^2)a^3}{Eh^3}\,,~~~\zeta=ak +2\pi n\,,\eeq{fff}
where the $A$-term corresponds to the plate inertia.

The results of calculation of the dispersive relation for some values of $\hat{\Gk}$ for $A=30, B=15$ are presented in \fig{a3}. Note that lowering of the frequency with the decrease of the hinge stiffness, which occurs at shorter waves, is caused by the decrease of the effective bending stiffness of the plate.

\begin{figure}[h!] 

\vspace{-0mm}\hspace{-15mm}
\hspace{30mm}\includegraphics*[width=0.5\textwidth]{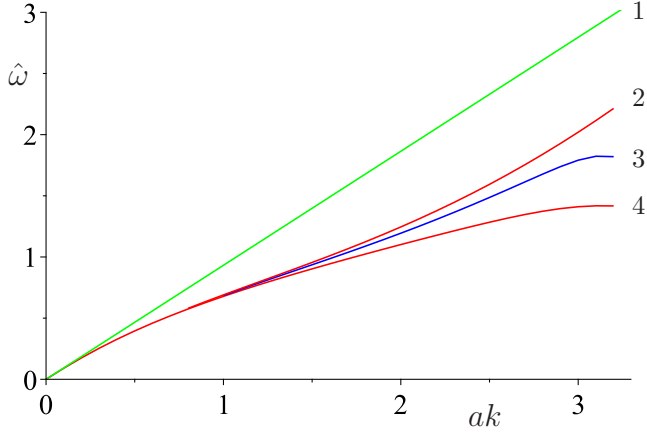}
\begin{picture}(0,0)(0,-100)
\put(-6, 55){{\footnotesize 1}}
\put(-6, 22){{\footnotesize 2}}
\put(-6, -1){{\footnotesize 3}}
\put(-6, -19){{\footnotesize 4}}
\put(-242, 30){$\hat{\Go}$}
\put(-68, -99){$ak$}

\end{picture}
\vspace{0mm}
\caption[]{The half-space elastic medium covered by the plate: the Rayleigh wave for the free surface) (1); the Rayleigh wave for the intact plate (with the sliding contact), $\Gk_M=\infty$ (2); the Rayleigh- Floquet wave for $\Gk_M=1/2$ (3) and for $\Gk_M=0$ (4).}
\label{a3}
\end{figure}

\section{Conclusions}
This paper expands the class of homogeneous system on segmented structures allowing thereby to treat (mathematically) structural discontinuities as singular strains and show how specific localized self-equilibrated loads produce the latter.

In some cases, the structural discontinuity breaks the continuity of internal forces, that is it acts as a non-self-equilibrate force.

Under this approach, there is no need for examining the system parts separately with subsequent conjugation. Only conditions related to the discontinuities remain to be satisfied, while the continuity in other respects preserves itself automatically.

In the analysis, the formulation discussed allows using the continuous Fourier transform as well as the discrete one. The coupled continuous and discrete transform arises under the uniform distribution of the discontinuities. The reader can see how a single discrete transform follows from the coupled one.

The technique is demonstrated by examining problems for the segmented flexural beam and plate interacting with some continual structures: the Floquet dispersion dependencies for waves in the flexural beam on Winkler's foundation, the Floquet gravity wave in water covered by a flexural plate, and the Floquet-Rayleigh wave in an elastic half-space covered by the plate.

The above partial problems for the plate interacting with some media are briefly considered here only as the examples of the technique used. These problems deserve to be studied thoroughly.

\section{Appendix} {\bf {\Large Wave equations for elastic body with discontinuities}
}

\vspace{3mm}\noindent
The wave equations (which can be seen, e.g., in Slepyan, 2002, p. 103) is
\beq \GD \Gf -\f{1}{c_1^2}\f{\p^2\Gf}{\p t^2} = - \f{f}{\Gl + 2\Gm}\,,~~~c_1= \sqrt{\f{\Gl + 2\Gm}{\Gr}}\,,\n
\GD \bfm{\psi} -\f{1}{c_2^2}\f{\p^2 \bfm{\psi}}{\p t^2} = - \f{ \bfm{p}}{\Gm}\,,~~~c_2= \sqrt{\f{\Gm}{\Gr}}\,,\eeq{we1}
where $\Gf$, $\bfm{\psi}$ and $f$,$ \bfm{p}$ are the (scalar and vector) potentials of the displacement $\bfm{u}$ and body force $\bfm{q}$, \res, such that
\beq \bfm{u} = \bfm{\nabla} \Gf +\bfm{\nabla} \wedge \bfm{\psi}\,,~~~\bfm{q} =  \bfm{\nabla} f +\bfm{\nabla} \wedge \bfm{p} \,.\eeq{we2}

We now consider the (fundamental) $x,y$ strain plane problem. The task is to determine the potentials $f$ and $\bfm{p}$ for the point-localized generalized forces corresponding to the delta-discontinuity. The latter places at the same point as the discontinuity (here it is $x=y=0$), where no localized mass is assumed. It follows that the force-discontinuity relations are the same as in statics, where the forces (clearly presented in Slepyan, 1981/1990) depend on time as a parameter. So we have no need showing them explicitly.
For mode I discontinuity $u_y(x,\pm 0) = \pm \Gd(x)$ (with the shear stresses $\Gs_{xy}(x,0)=0$) the forces are
\beq q_x = 2\Gm\f{1-4\nu}{3-4\nu}\Gd'(x)\Gd(y) = \f{\p f}{\p x} +\f{\p p}{\p y}\,,\n
q_y = 2\Gm\f{5-4\nu}{3-4\nu}\Gd(x)\Gd'(y) = \f{\p f}{\p y} -\f{\p p}{\p x}\,,\eeq{we3}
where $p=p_z$ and $\nu$ is Poisson's ratio. Note that only the normal-to-the-plane components of $\bfm{\psi}$ and $\bfm{p}$ remain  in the plane problem.

Using the (generalized) Fourier transform it can be found from these equations that
\beq f^F = \f{2\Gm}{k_x^2+k_y^2}\gl(\f{1-4\nu}{3-4\nu}k_x^2 +\f{5-4\nu}{3-4\nu}k_y^2 \gr)\,,\n
p^F=-\f{8\Gm}{3-4\nu}\f{k_x k_y}{k_x^2+k_y^2}\,.\eeq{we4}
It follows that
\beq f = 2\Gm \Gd(x)\Gd(y) -\f{4\Gm}{\pi(3-4\nu)}\f{x^2-y^2}{(x^2+y^2)^2}\,,\n
 p = - \f{8\Gm}{\pi(3-4\nu)}\f{xy}{(x^2+y^2)^2}\,,\eeq{we5}
where the expressions  apart from the delta-term are also singular at $x=y=0$. To show all the singularities explicitly, we rearrange the $f^F$ expression as follows
\beq f^F = f^F_{11}+f^F_{12}\,,~~f^F_{11}=2\Gm\f{1-4\nu}{3-4\nu}\,,~~ f^F_{12}=\f{8\Gm}{3-4\nu}\f{k_y^2}{k_x^2+k_y^2}\,,\n
f^F = f^F_{21}+f^F_{22}\,,~~f^F_{21}=2\Gm\f{5-4\nu}{3-4\nu}\,,~~ f^F_{22}=-\f{8\Gm}{3-4\nu}\f{k_x^2}{k_x^2+k_y^2}\,.\eeq{we5a}
The original functions
\beq f_{11} = 2\Gm\f{1-4\nu}{3-4\nu}\Gd(x)\Gd(y)\,,~~~ f_{21} = 2\Gm\f{5-4\nu}{3-4\nu}\Gd(x)\Gd(y)\,,\eeq{we5b}
fully define the forces \eq{we3}\,, \res, whereas the other terms $f^F_{12}$ and $\psi$, and  $f^F_{22}$ and $\psi$ cancel out
\beq \f{\p f_{12}}{\p x} +  \f{\p \psi}{\p y}=0~~~(-\I k_x f^F_{12} =  \I k_y\psi)\,,\n
 \f{\p f_{22}}{\p y} -  \f{\p \psi}{\p x}=0~~~(-\I k_y f^F_{22} =  -\I k_x\psi)\,.\eeq{we5c}

For mode II with the jump $u_x(x,\pm 0)=\pm \Gd(x)$ (under the condition $\Gs_{yy}(x,0)=0$), based on the relations as
\beq q_x= 2\Gm\Gd(x)\Gd'(y) =   \f{\p f}{\p x} +\f{\p p}{\p y}\,,\n
q_y = 2\Gm\Gd'(x)\Gd(y) =   \f{\p f}{\p y} -\f{\p p}{\p x}\,,\eeq{we6}
we find
\beq f^F = \f{4\Gm k_xk_y}{k_x^2+k_y^2}\,,~~~
 p^F=\f{2\Gm(k_x^2-k_y^2)}{k_x^2+k_y^2}\,,\eeq{we7}
and outside of $x=y= 0$, where the below expressions are singular
\beq f= \f{4\Gm}{\pi}\f{xy}{(x^2+y^2)^2}\,,~~~p = -\f{2\Gm}{\pi}\f{x^2-y^2}{(x^2+y^2)^2}\,.\eeq{we8}
Similarly to mode I, we can represent
\beq p^F= 2\Gm - \f{4\Gm k_y^2}{k_x^2+k_y^2} = -2\Gm + \f{4\Gm k_x^2}{k_x^2+k_y^2}\,.\eeq{we8a}
Now the first terms of the potential $p$ represent the forces $q_x$ and $q_y$, \res, whereas the other terms annihilate together with $f$.

Thus, with the above remarks, the force potentials are found as in \eq{we4} and \eq{we5} (mode I) and  \eq{we7} and \eq{we8} (mode II).

Finally, for mode III we have a single equation
\beq \f{\p^2 u_z}{\p x^2}+ \f{\p^2 u_z}{\p y^2}-\f{1}{c_2^2}\f{\p^2 u_z}{\p t^2} = - \f{q_z}{\Gm} = 2\Gd(x)\Gd'(y)\,.\eeq{we9}

\vspace{5mm}\noindent
{\bf Acknowledgement.} I am grateful to Professors John P. Dempsey and Hayley Shen for prompting me to consider this problem, and Professor Gennady Mishuris for his help.

\vspace{10mm}
\vskip 18pt
\begin{center}
{\bf  References}
\end{center}
\vskip 3pt

\inh Achenbach, J. D., and Epstein, H. I.,  1967. Dynamic interaction of a layer and a half-space. J. Eng. Mech. Div., Proc. Amer. Soc. Civ. Eng., 93, No. EM5, 24-42.

\inh Bremermann, H., 1965. Distributions, Complex Variables, and Fourier Transforms. Addison-Wesley Publishing Company. Reading, Massachusetts.

\inh Dempsey, J. P., Adamson, R. M., and DeFranco, S. J., 1995. Fracture of base-edge-cracked reverse-tapered plates. Int. J. Fract. 69, 281-294.

\inh Gueorguiev, Gregory McDaniel, J., Dupont, P., and Felsen, L. B., 2000. Analysis of Floquet wave generation and propagation in a plate with multiple arrays of line attachments. Journal of Sound and Vibration, 234(5), 819-840.

\inh Joglekar, D.M., Mitran, M., 2016. Analysis of flexural wave propagation through beams with a breathing crack using wavelet spectral finite element method. Mechanical Systems and Signal Processing 76-77, 576–591.

\inh Lighthill, J., 1978. Waves in Fluids. Cambridge University Press, Cambridge.

\inh Ozisik, M., and Akbarov, S. D., 2003. Rayleigh wave propagation in a half-plane covered with a prestressed layer under complete and incomplete interfacial contact. Mechanics of Composite Materials, 39, No. 2, 177-182.

\inh Rice, J. R., and Levy, N., 1972. The part-through surface crack in an elastic plate. J. Appl. Mech. 39, 185-194.

\inh Slepyan, L. Dempsey, J. and Shekhtman, I., 1995. Asymptotic
Solutions for Crack Closure in an Elastic Plate under Combined
Extension and Bending. J. Mech. Phys. Solids, 43, No. 11, 1727-1749.

\inh Slepyan, L.I. 1981/1990. Mechanics of Cracks. Sudostroenie,
Leningrad (in Russian. 1st addition: 1981, 2nd addition: 1990).

\inh Slepyan, L.I., 2002. Models and Phenomena in Fracture
Mechanics. Springer, Berlin.

\end{document}